\documentclass[manuscript,screen]{acmart}
\graphicspath{{Images/}}
\AtBeginDocument{%
  \providecommand\BibTeX{{%
    \normalfont B\kern-0.5em{\scshape i\kern-0.25em b}\kern-0.8em\TeX}}}

\usepackage{multirow}
\usepackage{natbib}
\acmConference[CHI'21]{ACM SIGCHI Conference on Human Factors in Computing Systems 2021}{May 8--13, 2021}{Online Virtual Conference}
\acmBooktitle{ACM SIGCHI Conference on Human Factors in Computing Systems 2021, May 8--13, 2021, Online Virtual Conference}
\acmPrice{15.00}
\acmISBN{978-1-4503-XXXX-X/18/06}
\usepackage[utf8]{inputenc}
\usepackage{multirow}
\begin{document}

\title{\textit{WAccess} - A Web Accessibility Tool based on WCAG 2.2, 2.1 and 2.0 Guidelines}

\author{Kowndinya Boyalakuntla, Akhila Sri Manasa Venigalla AND Sridhar Chimalakonda}

\email{\{cs17b032,cs19d504,ch\}@iittp.ac.in}

\affiliation{%
  \institution{\\
  \textit{Research in Intelligent Software \& Human Analytics (RISHA) Lab}\\ 
  Department of Computer Science \& Engineering\\
  Indian Institute of Technology Tirupati}
  \city{Tirupati}
  \country{India}}
  
\settopmatter{printacmref=false}
\setcopyright{none}
\renewcommand\footnotetextcopyrightpermission[1]{}
\pagestyle{plain}
\begin{abstract}
The vision of providing access to all web content equally for all users makes web accessibility a fundamental goal of today’s internet. Web accessibility is the practice of removing barriers from websites that could hinder functionality for users with various disabilities. Web accessibility is measured against the accessibility guidelines such as WCAG, GIGW, and so on. WCAG 2.2 is the latest set of guidelines for web accessibility that helps in making websites accessible.  The web accessibility tools available in the World Wide Web Consortium (W3C), only conform up to WCAG 2.1 guidelines, while no tools exist for the latest set of guidelines. Despite the availability of several tools to check the conformity of websites with WCAG 2.1 guidelines, there is a scarcity of tools that are both open source and scalable. To support automated accessibility evaluation of numerous websites against WCAG 2.2, 2.1, and 2.0 we present a tool, \textit{WAccess}. \textit{WAccess} highlights violations of 13 guidelines from WCAG 2.0, 9 guidelines from WCAG 2.1, and 7 guidelines from WCAG 2.2 of a specific web page on the web console and suggests the fix for violations while specifying violating code snippet simultaneously. We evaluated \textit{WAccess} against 2227 government websites of India and observed a total of about 6.1 million violations.  
\end{abstract}



\maketitle

\section{INTRODUCTION}
Usability and accessibility are the commonly used terms in the context of enhancing the user experience for users of the world wide web~\cite{hanson2013progress}. 
While there is no universally accepted view or definition for usability, a commonly accepted definition presented in ISO 9241-11 standard explains that a product is considered to be useful if the specified users can accomplish specified tasks effectively, efficiently, and with satisfaction \cite{iso1998ergonomic}. 
Currently, most public and private activities heavily rely on web-based services, making it critical for the web to be usable on multiple devices such as desktops, tablets and mobile phones \cite{jeong2020dynamic}, by any individual, irrespective of any physical or mental barriers \cite{sullivan2000barriers}. 
Usability of websites is   influenced by the level of accessibility of websites \cite{giraud2018web}. 
Services that involve promoting businesses, training defense systems and so on using recommender systems also rely on the web portals for crowd sourced information such as identifying areas of interest using user eye tracking, user keyboard movement and so on \cite{eraslan2020best, mazumdar2020cold, flores2021utilizing, vidyapu2020investigating}. An accessible portal could thus be capable of collecting more information from a wide and diverse range of individuals.   
Web accessibility has attracted significant attention from researchers and governments across the globe since its inception to provide better access to the websites \cite{rau2016evaluation}. 
Researchers have emphasized the need for web developers to abide by accessibility guidelines of websites to support broader usability of web \cite{lazar2004improving}. Moreover, many websites are being designed based on the templates, thus requiring care to be taken to incorporate accessibility principles in designing these templates \cite{alarte2019web}.

User satisfaction and consequently user experience on visiting the website is considered as an important criteria while designing a website \cite{alexander2021influence}. Many attempts such as approaches to improve the trustworthiness of the websites, protect visitors of the website, designing culturally adapted websites and so on are being made to increase the usability and usage of the websites  \cite{carpineto2020experimental, alexander2021influence, roy2021integrated}.
Despite the massive digitization, a significant amount of web content and e-services are not accessible to a large section of users today~\cite{hanson2013progress}. 
When it comes to utilizing the web, a diverse range of users exists, including visually impaired or disabled groups and older adults, who might find it difficult to read through a page \cite{fritz2019customization}. 
Providing content in a similar fashion to all types of user groups as presented to user groups without disabilities makes the web inaccessible to the rest of these groups \cite{carvalho2018accessibility}.
While improving reliability and trustworthiness of a website \cite{carpineto2020experimental, roy2021integrated} and presenting culturally relevant elements on the page \cite{alexander2021influence} are some important aspects to ensure customer safety and satisfaction, making the website accessible to diverse user groups and thus providing better user experience to a diverse range of users is also essential to increase the reach of the website.   
Several attempts are being made to address this challenge of web accessibility \cite{moreno2019harmonization, crespo2016social4all, gay2010achecker}. 
In order to overcome these challenges, standard guidelines such as WCAG have been proposed to support website developers and designers to ensure website accessibility. 
The latest revised WCAG guidelines are WCAG 2.2\footnote{\url{https://www.w3.org/TR/WCAG22/}} proposed in the year 2021. 
Several tools such as Achecker \cite{gay2010achecker} and CAC \cite{klein2014checking} have been developed to evaluate websites against different versions of guidelines such as WCAG 1.0\footnote{\url{https://www.w3.org/TR/WAI-WEBCONTENT/}},  WCAG 2.0\footnote{\url{https://www.w3.org/TR/WCAG20/}} guidelines, Stanca Act\footnote{\url{https://www.levelaccess.com/accessibility-regulations/italy/}} (Italian accessibility guidelines), and so on. Most of the existing tools focus on highlighting errors based on WCAG 2.0 guidelines, while fewer tools exist for evaluating websites based on WCAG 2.1 guidelines \cite{klein2014checking}, and no tools exist to evaluate websites based on WCAG 2.2 guidelines. 
World wide web consortium (W3C) lists 132 web accessibility evaluation tools for WCAG 2.0, 67 evaluation tools for WCAG 2.1, while none are listed for WCAG 2.2.\footnote{\url{https://www.w3.org/WAI/ER/tools/}}

Of the 67 tools, only 15 are available as open-source to evaluate websites against WCAG 2.1 guidelines. 
Tools supporting command-line interfaces help verify the accessibility of a large number of websites automatically~\cite{fernandes2011web}. 
In contrast, tools designed as browser plugins help in an easy and quick understanding of the accessibility of the website. However, based on the above criteria, \textit{QualWeb}\footnote{\url{http://qualweb.di.fc.ul.pt/evaluator/about}} is the only tool listed to support both command-line and browser plugin facility. 
We observe that \textit{QualWeb} is not available as a browser extension yet. 
It also requires the further installation of other packages such as npm and revised chromium-browser to use the command line interface version, making it difficult to use the tool. 
Even after installing the required dependencies, many errors occurred, preventing the functioning of the tool\footnote{Snapshots of the errors occurred are presented in this document - {\url{https://osf.io/k9v8a/?view_only=9b7799ccf554412f9cdaafa61da4bf52}}}.
This indicates the need for better tools and approaches that could evaluate the accessibility of websites against WCAG 2.1 guidelines and consequently be used in web development. 

Hence, in this paper, we propose \textit{WAccess\footnote{\url{https://sites.google.com/iittp.ac.in/waccess}}}, a tool to assess web accessibility of websites against the latest WCAG 2.2 guidelines, along with WCAG 2.1 and 2.0 guidelines. 
\textit{WAccess} displays a \textit{list of errors} with respect to \textit{accessibility guidelines}, the \textit{code snippet} that causes the \textit{error} and a \textit{suggested fix}.
We call the collective set of guidelines considered for \textit{WAccess} as WCAG2 series. 
Indian Government websites contain vast information and are critical for good governance in the country \cite{paul2020accessibility}.
Since a large number of users are intending to use the government's e-services, especially in the Indian context, the massive volume of government information is incorporated onto the web \cite{ismail2016accessibility}. 
This aspect resulted in the growth of research on evaluating the accessibility of government websites \cite{ismail2018accessibility, mtebe2017accessibility, rau2016evaluation, karaim2019usability}. 
However, these evaluations are performed on a smaller number of websites ranging from 10 to 302 website evaluations, as the existing tools for evaluating web accessibility are complex in nature for performing guideline automated analysis. Also, these evaluations are based on WCAG 2.0 and WCAG 1.0, which were proposed prior to WCAG 2.1, in 2010. While \citet{narasimhan2012accessibility} have automatically evaluated accessibility of a larger number of GoI websites (7800) using \textit{Achecker\footnote{\url{https://achecker.achecks.ca/checker/index.php}}}, this evaluation is confined only to WCAG 2.0 guidelines, owing to the limitation of \textit{AChecker} in supporting evaluation only upto WCAG 2.0.

% \color{blue} As an attempt towards analyzing accessibility of multiple government websites against WCAG2 series, and simultaneously towards evaluating the usefulness of \textit{WAccess}, we performed a study on the accessibility of 2246 Indian government websites using \textit{WAccess}. 
As an attempt towards analyzing accessibility of multiple government websites against WCAG2 series, we performed a study on the accessibility of 2227 Indian government websites using \textit{WAccess}, and through this study, \textit{WAccess} could detect 6131929  violations. The results of the study are presented here\footnote{\url{https://osf.io/fnx4t/?view_only=1769872b5dd447fbbb30fe47ecf88ece}}. Source code of \textit{WAccess} is available at: \url{https://github.com/Kowndinya2000/WAccess}.  

The contributions of this paper are:
\begin{enumerate}
    \item \textit{WAccess}, a tool to check the conformance of websites against web accessibility standards specified by W3C, in selected guidelines of the latest WCAG 2.2, WCAG 2.1 and WCAG 2.0. 
    \item Accessibility study on 2227 Indian government websites present in the Government of India web directory using \textit{WAccess} tool.
    \item A list of violations for each of the websites considered for the accessibility study.  
\end{enumerate}

% \color{red}
The rest of the paper is organized as follows: we describe the related work in Section \ref{sec:related_work}. We then discuss the design of \textit{WAccess} in Section \ref{sec:design}. Section \ref{sec:eval} presents evaluation of Indian government websites by \textit{WAccess}, followed by discussion and limitations in Section \ref{sec:discussion}. We conclude the paper and present some future work in Section \ref{sec:conclusion}.
% \color{black}

\section{RELATED WORK}
\label{sec:related_work}
Web Accessibility is considered as an important issue in the current digital world, leading to the emergence of  several approaches and guidelines to improve accessibility \cite{moreno2019harmonization, crespo2016social4all, gay2010achecker, broccia2020flexible}. Moreno et al. have emphasized the need for standardizing the web accessibility standards across the world, and suggested the use of WCAG guidelines \cite{moreno2019harmonization}.

There are several tools and techniques in the literature to evaluate websites' accessibility against WCAG guidelines \cite{gay2010achecker, klein2014checking, takata2004accessibility}. 
Takata et al. \cite{takata2004accessibility} proposed a tool to verify the accessibility and syntactic correctness of a website. The tool supports verification of any XML document, by separating out the guidelines to facilitate easy modification of the guidelines \cite{takata2004accessibility}. \textit{Achecker} \cite{gay2010achecker} is a standalone open-source tool to analyze the extent to which a website adheres to a set of accessibility guidelines. It facilitates users to choose the desired accessibility guidelines, which are to be evaluated for a website from a pre-loaded list \cite{gay2010achecker}. \textit{WAVE} tool has been proposed to identify accessibility errors with respect to WCAG guidelines, to support web developers in developing web pages that are accessible to all, irrespective of individuals with disabilities\footnote{\url{http://wave.webaim.org/}}. \textit{CAC} evaluates a website against WCAG 2.0 guidelines for accessibility issues \cite{klein2014checking}. \textit{CAC} also reports issues to the users by highlighting them on the webpage and proposes possible solutions to resolve the issues \cite{klein2014checking}. Crespo et al. also suggest a novel approach to support the rectification of a few accessibility issues in websites, based on evaluating adherence to a set of accessibility guidelines \cite{crespo2016social4all}. Broccia et al. have highlighted the need for tools and approaches to validate accessibility of websites and presented a tool to support WCAG 2.1 guidelines \cite{broccia2020flexible}. They further used the presented tool to evaluate twelve websites in six different categories including health, public bodies, travel and so on \cite{broccia2020flexible}.

Web usability evaluation of government websites has been performed by several countries such as China \cite{rau2016evaluation}, Tanzania \cite{mtebe2017accessibility}, Kyrgyz Republic \cite{ismailova2017web}, India \cite{katre2011expert} and researchers have observed that most of the websites fail to meet the minimal accessibility standards. Recently, Spina has analyzed WCAG 2.1 guidelines for libraries and found that it is important to update the existing tools to support WCAG 2.1 \cite{spina2019wcag}.

There is scarce research on web accessibility in the Indian context. Researchers have evaluated web accessibility of banking websites \cite{kaur2014banking}, educational institutions~\cite{ismail2018accessibility} among other websites. An expert manual study of 28 Government of India (GoI) websites found that most of the websites are either down or have accessibility issues in the year 2011 \cite{katre2011expert}. 
A recent case study was performed on 164 Indian government ministry websites to assess usability, accessibility and mobile readiness with the help of an automatic tool named TAW\footnote{\url{https://www.tawdis.net/}}, however the evaluation is based on WCAG 2.0 guidelines only~\cite{agrawal2021assessing}.
A study performed by the Center for Internet and Society on 7800 websites of GoI using existing web accessibility evaluation tools based on WCAG 2.0 found an average of 63 errors per home page, with a few pages crossing 1000 errors~\cite{narasimhan2012accessibility}. 
In a study conducted to assess the accessibility of 15 GoI portals concerning WCAG 2.0 (2008) and GIGW Guidelines, \citet{patra2018accessibility} have listed specific aspects that are to be considered to improve accessibility websites \cite{patra2018accessibility}. 
Recently,~\citet{paul2020accessibility} studied the accessibility and usability of 65 Indian government websites against WCAG 2.0 and WCAG 1.0 using automated tools specified for these guidelines \cite{paul2020accessibility}. 
The results revealed that the considered websites do not prioritize accessibility aspect, eventually leading to the low quality of government websites in India~\cite{paul2020accessibility}.  

As mentioned above, majority of the existing approaches and tools for WCAG 2.1 are proprietary. Only 13 tools are open-source, out of which only \textit{QualWeb} works both as a browser plugin and a command line interface. 
However, \textit{QualWeb} did not work when attempted to run and also has the overhead of installing other packages as pre-requisites to run.
Furthermore, no tools exist for the latest WCAG 2.2 guidelines. 
Hence, we propose \textit{WAccess}, as a Google Chrome plugin that aims to verify the accessibility of websites based on WCAG2 series. We evaluated 2227 Indian government websites with \textit{WAccess}.

\section{DESIGN AND DEVELOPMENT OF \textit{WAccess}}
\label{sec:design}
\subsection{WEB ACCESSIBILITY GUIDELINES}
The Web Accessibility Initiative (WAI) of the W3C has proposed several standards with the goal of “access to the Web by everyone,”  with each of the standards having a layered set of principles, guidelines, success criteria, and sufficient and advisory techniques. 

WCAG 1.0 is one of the initial web accessibility guidelines that appeared in 1999 with a revised version of WCAG 2.0, 2.1 in 2008 and 2018 respectively and the recent standard version, WCAG 2.2, in 2021. 

Each of these standards is backward compatible with each other. Since the start of WCAG 2.0, three conformance levels of accessibility denoted as A (basic accessibility), AA (desirable accessibility), AAA (full accessibility) are introduced that could be customized as per specific needs of the web content and web content providers. In addition to WCAG, two more standards - Authoring Tool Accessibility Guidelines (ATAG) 2.0 and User Agent Accessibility Guidelines (UAAG) 2.0, are proposed to assist web developers and users with disabilities by enhancing user agents in the websites, such as text-to-speech support. 
\begin{figure}[h]
    \centering
    \includegraphics[height = 7cm, width = \linewidth]{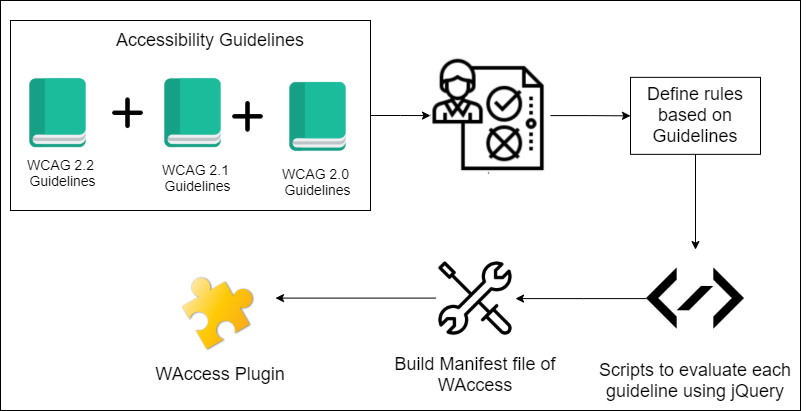}
    \caption{Design Methodology of \textit{WAccess}}
    \label{fig:approach}
\end{figure}
% \begin{figure}[h]
%     \centering
%     \includegraphics[height = 7cm, width = \linewidth]{Images/WAccess-Overview-grey-scale.png}
%     \caption{Design Methodology of \textit{WAccess}}
%     \label{fig:approach2}
% \end{figure}
\subsection{DESIGN METHODOLOGY}
\textit{WAccess} is implemented as a browser plugin, when activated, displays accessibility violations on the web console.
By analyzing the guideline definitions and success criteria, we framed rules and formulated the necessary conditions for each guideline in individual JavaScript files. We narrowed down the scope of accessibility testing to the extent we can automate by considering part of the success criterion of guidelines requiring human investigation. For instance, (1) Guideline 2.4.4 requires the purpose of a link to be effectively described in its linked text (2) According to Guideline 1.1.1, providing null alt text will help assistive technology to ignore decorative images, but whether image is decorative or not is subjective,   
% or
% if an image's textual description is precise  \color{black}
% determining the relevance of alternate text presented for images, to the image content on the webpage, 
 are not in the scope of \textit{WAccess}. The above described situations will require human observation to make a decision on whether the guideline has been violated or not.

Violations highlighted by \textit{WAccess} would form a subset of the all possible violations, thus implying if \textit{WAccess} highlights zero violations for a guideline then it does not mean the success criterion is completely met but it means no violations were spotted in the implementation scope of \textit{WAccess}.  
\textit{WAccess} plugin has been designed based on the approach shown in Fig \ref{fig:approach}. 

\emph{WAccess} was developed to help determine if web content meets accessibility standards with WCAG 2.2, 2.1 and 2.0 in consideration. 

We focus on WCAG 2.0, 2.1 and 2.2 which are based on four core principles and 29 guidelines with each guideline having multiple success criteria. 
The four core principles are as follows: 
\begin{itemize}
    \item\textbf{Perceivable:} Users must be able to perceive all relevant information in your content.
    \item\textbf{Operable:} Users must be able to operate the interface successfully.
    \item \textbf{Understandable:} Users must be able to understand the information and operation of the interface.
    \item \textbf{Robust:} Content must be accessible to all users and should be easily interpreted by wider range of user agents.
\end{itemize}

%  only wcag 2.2 - 34,837 violations
% only wcag 2.1  - 2,97,606

\begin{table}[]
\begin{tabular}{|l|l|l|l|l|l|}
\hline
Principle                 & ID     & Description                 & Level    & Total Violations    & Websites \\ \hline
\multirow{5}{*}{Operable} & 2.4.11 & Focus Appearance (Minimum)  & AA       & 1393698             & 2227     \\ \cline{2-6} 
                          & 2.4.12 & Focus Appearance (Enhanced) & AAA       & 1654014            & 2227     \\ \cline{2-6}  
                          & 2.4.13 & Page Break Navigation       & A         & 2142               & 292     \\ \cline{2-6}  
                          & 2.5.7  & Dragging Movements          & AA        & 346825             & 2102     \\ \cline{2-6}  
                          & 2.5.8  & Target Size (Minimum)       & AA        & 1001991            & 2221     \\ \hline
\multirow{2}{*}{Understandable}            & 3.2.7  & Visible Controls            & A        & 0                & 0     \\ \cline{2-6}
                          & 3.3.7  & Accessible Authentication   & A        & 9358                & 1458     \\ \hline
\end{tabular}
\caption{7 WCAG 2.2 Guidelines considered for developing \textit{WAccess} with total violations observed and number of websites violating each guideline}
    \label{tab:table_2.2}
\end{table}

\begin{table}[]
\begin{tabular}{|l|l|l|l|l|l|}
\hline
Principle                    & ID     & Description                 & Level    & Total Violations    & Websites \\ \hline
\multirow{4}{*}{Perceivable} & 1.3.5  & Identify Input Purpose      & AA       & 7991                & 1303    \\ \cline{2-6} 
                             & 1.3.6  & Identify Purpose            & AAA      & 3899                & 923   \\ \cline{2-6} 
                             & 1.4.11 & Non-text Contrast           & AA       & 40558               & 1547    \\ \cline{2-6} 
                             & 1.4.13 & Content on Hover or Focus   & AA       & 0                   & 0    \\ \hline
\multirow{4}{*}{Operable}    & 2.1.4  & Character Key Shortcuts     & A        & 0                   & 0     \\ \cline{2-6} 
                             & 2.3.3  & Animation from Interactions & AAA      & 0                   & 0   \\ \cline{2-6} 
                             & 2.5.3  & Label in Name               & A        & 62923               & 1782     \\ \cline{2-6} 
                             & 2.5.5  & Target Size                 & AAA      & 56497               & 1823   \\ \hline
Robust                       & 4.1.3  & Status Messages             & AA       & 0                   & 0    \\ \hline
\end{tabular}
\caption{9 WCAG 2.1 Guidelines considered for developing \textit{WAccess} with total violations observed and number of websites violating each guideline}
    \label{tab:table_2.1}
\end{table}

\begin{table}[]
\begin{tabular}{|l|l|l|l|l|l|}
\hline
Principle                       & ID    & Description               & Level    & Total Violations    & Websites \\ \hline
\multirow{6}{*}{Perceivable}    & 1.1.1 & Non-text Content          & A        & 93040               & 1996     \\ \cline{2-6} 
                                & 1.3.1 & Info and Relationships    & A        & 3364                & 850     \\ \cline{2-6} 
                                & 1.4.1 & Use of Color              & A        & 0                   & 0     \\ \cline{2-6} 
                                & 1.4.3 & Contrast (Minimum)        & AA       & 469170              & 2168     \\ \cline{2-6} 
                                & 1.4.4 & Resize text               & AA       & 32545               & 1448     \\ \cline{2-6} 
                                & 1.4.6 & Contrast (Enhanced)       & AAA      & 521437              & 2170     \\ \hline
\multirow{4}{*}{Operable}       & 2.1.1 & Keyboard                  & A        & 0                   & 0     \\ \cline{2-6}
                                & 2.2.2 & Pause, Stop, Hide         & A        & 1283                & 834     \\ \cline{2-6} 
                                & 2.4.4 & Link Purpose (In Context) & A        & 230606              & 2049     \\ \cline{2-6} 
                                & 2.4.6 & Headings and Labels       & AA       & 2208                & 2208     \\ \hline
\multirow{2}{*}{Understandable} & 3.1.1 & Language of Page          & A        & 2187                & 2187     \\ \cline{2-6} 
                                & 3.3.2 & Labels or Instructions    & A        & 193624              & 1678     \\ \hline
Robust                          & 4.1.1 & Parsing                   & A        & 2569                & 584     \\ \hline
\end{tabular}
\caption{13 WCAG 2.0 Guidelines considered for developing \textit{WAccess} with total violations observed and number of websites violating each guideline}
    \label{tab:table_2.0}
\end{table}
% 3.3.2 guideline results from part1 or part2

\textit{WAccess} considers 7 WCAG 2.2 guidelines, 9 WCAG 2.1 guidelines, 13 WCAG 2.0 guidelines, which do not require human intervention, as shown in Tables \ref{tab:table_2.2}, \ref{tab:table_2.1}, and \ref{tab:table_2.0} to evaluate the accessibility of a website. 
We integrate the guidelines into 4 classes -(i) \textit{aria-related}, (ii) \textit{color-contrast related}, (iii) \textit{HTML-check related} and (iv) \textit{interaction-related}.
These guidelines are reviewed to identify and sort the best practices required to meet the criteria of all guidelines. 
Based on the best practices observed, rules are defined to evaluate a web page against the specified criteria. 
Scripts to check the accessibility of a website based on the rules defined are written using JQuery. 
Each of these scripts are designed to address one accessibility guideline. A manifest.json file is built to run the all these guideline specific javascript files.

\section{Evaluation and Results}
\label{sec:eval}
% \textit{WAccess} performs accessibility check pointing automatically, we chose all the websites from GOI web directory and performed the analysis. 
% On each website in the evaluation list, \textit{WAccess} evaluates the code violations in HTML markup loaded on the website and sends the violations to a database, that can be exported to various formats such as .json, .csv for further analysis. 
% With \textit{WAccess} we find number of violations conforming to 7 of WCAG 2.2 and 9 of WCAG 2.1 guidelines, and each violation is described with the code snippet causing violation and a possible correction to avoid the violation. 
% \emph{WAccess} is evaluated on 2246 government websites, and a graph highlighting the distribution
% violations across different guideline is shown in Fig. \ref{fig:eval_res2}.

\begin{figure}
    \centering
    \includegraphics[scale=0.6]{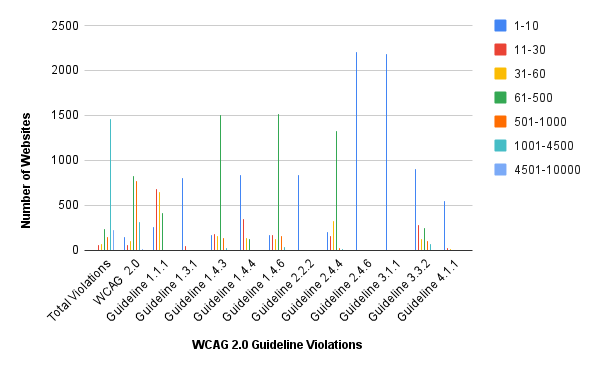}
    \caption{Distribution of violations with respect to WCAG 2.0 guidelines}
    \label{fig:wcag_2_0}
\end{figure}

\begin{figure}
    \centering
    \includegraphics[width = \linewidth]{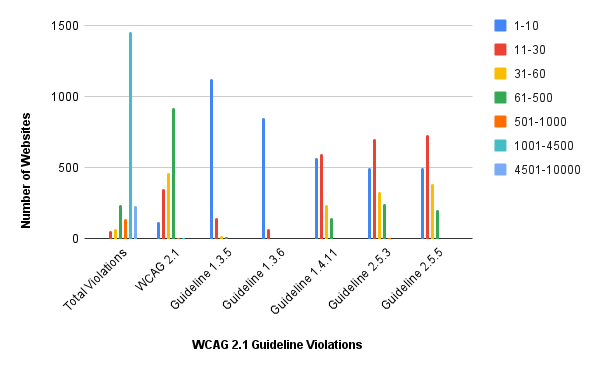}
    \caption{Distribution of violations with respect to WCAG 2.1 guidelines}
    \label{fig:wcag_2_1}
\end{figure}

\begin{figure}
    \centering
    \includegraphics[width = \linewidth]{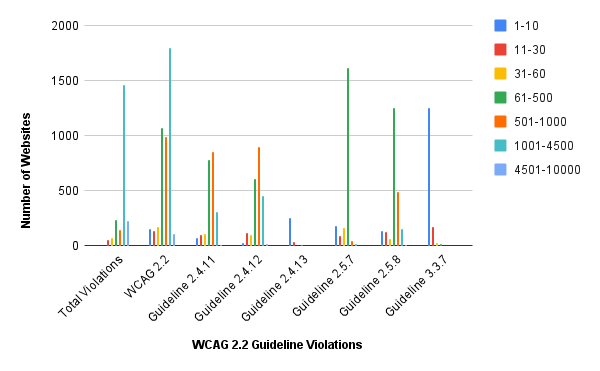}
    \caption{Distribution of violations with respect to WCAG 2.2 guidelines}
    \label{fig:wcag_2_2}
\end{figure}

\begin{figure}
    \centering
    \includegraphics[width = \linewidth]{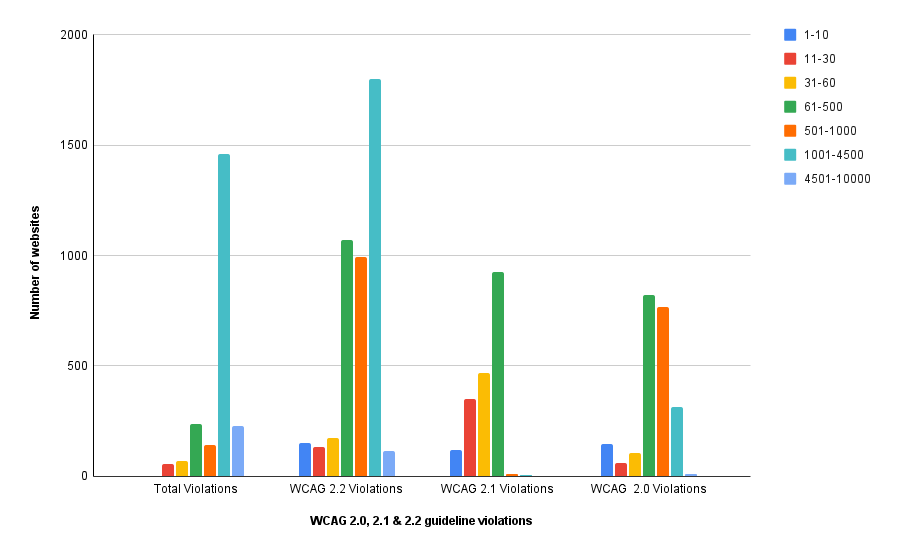}
    \caption{Distribution of violations across WCAG guideline series}
    \label{fig:wcag_total}
\end{figure}

Through Digital India, government of India made several of its services and governance available through digital platforms that is accessible to all the citizens of the country.
Considering the frequent and wide use of government of India websites, we chose to perform a case study on these websites to investigate the accessibility aspect. GOI web directory is a one source guide to all the government websites of India. We navigated to each state directory and downloaded the PDFs containing URLs of the e-websites and curated them by extracting the links from all the PDFs collected. Of the 3995 websites collected we found 2227 to be actively working and we carried out the analysis on them and further reported the results. For each website we stored the number of violations for each guideline and respective error and fix messages along with code snippet. We consolidated number of violations into different categories and plotted different graphs for each of the WCAG2 series (2.0, 2.1 and 2.2) in Fig. \ref{fig:wcag_2_0}, Fig. \ref{fig:wcag_2_1} and Fig. \ref{fig:wcag_2_2} respectively and violation distribution across each of the three WCAG guideline series in Fig. \ref{fig:wcag_total}.

\begin{table}[]
\begin{tabular}{|c|c|c|c|}
\hline
\multirow{2}{*}{Guideline} & \multicolumn{2}{c|}{Number of Violations} & \multirow{2}{*}{Conformance Level} \\ \cline{2-3}
                           & Aadhaar             & Commerce            &                                    \\ \hline
1.1.1                      & 57                  & 202                 & \multirow{7}{*}{A}                 \\ \cline{1-3}
1.3.1                      & 3                   & 2                   &                                    \\ \cline{1-3}
2.4.4                      & 324                 & 546                 &                                    \\ \cline{1-3}
2.4.13                     & 0                   & 22                  &                                    \\ \cline{1-3}
2.5.3                      & 138                 & 77                  &                                    \\ \cline{1-3}
3.1.1                      & 1                   & 1                   &                                    \\ \cline{1-3}
3.3.7                      & 6                   & 1                   &                                    \\ \hline
1.3.5                      & 5                   & 1                   & \multirow{8}{*}{AA}                \\ \cline{1-3}
1.4.3                      & 1085                & 833                 &                                    \\ \cline{1-3}
1.4.4                      & 177                 & 7                   &                                    \\ \cline{1-3}
1.4.11                     & 5                   & 48                  &                                    \\ \cline{1-3}
2.4.6                      & 1                   & 1                   &                                    \\ \cline{1-3}
2.4.11                     & 2404                & 2417                &                                    \\ \cline{1-3}
2.5.7                      & 398                 & 757                 &                                    \\ \cline{1-3}
2.5.8                      & 1967                & 2101                &                                    \\ \hline
1.3.6                      & 22                  & 15                  & \multirow{5}{*}{AAA}               \\ \cline{1-3}
1.4.6                      & 1175                & 1201                &                                    \\ \cline{1-3}
2.4.12                     & 2513                & 1709                &                                    \\ \cline{1-3}
2.5.5                      & 58                  & 118                 &                                    \\ \cline{1-3}
Total                      & 10339               & 10059               &                                    \\ \hline
\end{tabular}
\caption{Violations observed for UIDAI Aadhaar and Commerce Website}
\label{tab:case_study}
\end{table}

% \begin{figure}
%     \centering
%     \includegraphics[width = \linewidth]{Images/chart-waccess.png}
%     \caption{Results of study of 2246 Indian Government Websites using \textit{WAccess} for 9 WCAG 2.1 guidelines, and 7 WCAG 2.2 guidelines}
%     \label{fig:eval_res2}
% \end{figure}


% \begin{figure}
%     \centering
%     \includegraphics[width = \linewidth]{Images/chart_wcag.png}
%     \caption{Distribution of violations across all guidelines}
%     \label{fig:wcag_total}
% \end{figure}

\subsection{Guidelines and Violations}
In this section, we describe the violations resulted from the study on the chosen government websites.
We tabulated the violations observed for each guideline and number of websites violating each guideline for the three WCAG collections separately in Tables \ref{tab:table_2.2}, \ref{tab:table_2.1}, and \ref{tab:table_2.0}. 
We base our observations from the results obtained in Fig. \ref{fig:wcag_2_0}, Fig. \ref{fig:wcag_2_1}, Fig. \ref{fig:wcag_2_2}, Fig. \ref{fig:wcag_total}, and Tables \ref{tab:table_2.2}, \ref{tab:table_2.1}, \ref{tab:table_2.0}, and \ref{tab:case_study}. We observed 6.1 million violations across the 2227 websites, of which 4.57 million violations correspond to WCAG 2.2, 0.17 million to WCAG 2.1 and 1.55 million to WCAG 2.0 guidelines, as shown in the Fig. \ref{fig:wcag_total}.

\textit{Guideline 1.3.5 Identify Input Purpose (AA)}  requires the purpose for each input field to be defined in order to programmatically determine the information later. 
Ensuring this guideline criterion helps to retrieve information such as contact, billing information or login details automatically by the browser and this can benefit users with poor memory. 
Users with movement disorder can make use of the auto-fill feature by lessening the requirement for manual input in filling forms. 
This guideline has been violated at least once by 1303 among the 2227 websites chosen, with 161 being the maximum number of violations observed for a website\footnote{\url{http://wapcos.gov.in/}}.

\textit{1.3.6 Identify Purpose (AAA)} intents to guarantee that people interface components on the website can be programmatically determined.  This empowers user agents to decide the visual presentation of purpose of components there by making them more understandable for the user. 
Achieving the success criterion guarantees recognizing purpose behind UI elements and people with trouble in perusing or with intellectual inabilities (difficult to retain focus) would find this guideline useful. 
1304 websites have not violated this guideline, while it has been violated for a maximum of 55\footnote{\url{http://msmediagra.gov.in/aboutus.htm}} times among the websites chosen. 
% http://msmediagra.gov.in/aboutus.htm - 55 violations @1.3.6 - max 

\textit{Guideline 1.4.11 Non-Text Contrast (AA)} tells that low contrast controls are hard to see, and might be totally missed by individuals with a visual weakness, and hence the visual representation of UI components and graphical objects should have a contrast ratio of at least three to one against adjacent colors. The purpose of this success criterion is to ensure that active user interface components and meaningful graphics are recognizable by users with tolerably low vision without the need of additional assistive technologies. This guideline was violated at least once by 1547 websites, with a maximum at 391\footnote{\url{http://jkhighereducation.nic.in/order.html}} by any website. Less than 30 violations per website make 75\% of the non-zero violations. 

\textit{Guideline 2.5.3 Label in Name (A)} intends to guarantee that text which label a component visually should match the text associated with the component programmatically. This empowers sighted users who depend on screen readers as the text they hear apparently matches the text they see on the screen

% visible labels as a way to interact with the components. 
% states that text or images of text must contain accessibility name which 
% matches the text that is visually presented. 
% The success criterion intents to ensure that words which label a component visually match the words that label a component programmatically. 
% Mismatches create extra cognitive function for speech input users to remember a different speech command than the visual label of the control.

More than 90\% of the websites failed to meet this guideline. One among the 2227 websites chosen, violated this guideline for a maximum of 1478\footnote{\url{http://sknau.ac.in/}} times. Maximum number of websites violating this guideline fell in the violation range 11-30.  

\textit{Guideline 2.5.5 Target Size (AAA)} requires the controls to be large enough to see and touch. Users with low vision might experience issues in seeing a small target. Implementing the success criterion will assist individuals having difficulty with fine motor movements in activating a small target. Only 5.2\% of the websites do not violate this guideline. Maximum of 891\footnote{\url{http://www.tourism.rajasthan.gov.in/}} violations were observed for this guideline. 

\textit{Guideline 2.5.7 Dragging Movements (AA)} Success criterion of this guideline ensures dragging elements to not limit to single pointer access. Drag and drop actions require a reasonably precise dragging motion, hence  for users struggling with performing dragging cannot maintain contact in keeping their finger on the button or screen without accidentally releasing. 
Almost all the websites (94.3\%) found to be violating this guideline. Maximum number of violations referring to this guideline were ranged from 501-1000, with a maximum violated by any website at 9109.

\textit{Guideline 3.3.7 Accessible Authentication (A)} states that users with certain cognitive disabilities struggle with memorizing authentication credentials (such as usernames and passwords). This guideline requires at least one other authentication mechanism such as forget password link, and this would help people with memory related issues or perception-processing in-capacities to be able to authenticate without much of a cognitive function.
% requires that authentication must be possible with forget password links. People forget their passwords, and transcribing authentication codes sent to their phone into the page can be prone to error. 
Nearly, 34.5\% of the websites, found to be meeting the requirements referring to this guideline. Majority of the websites violated this guideline less than 10 times, with maximum number of violations at 161\footnote{\url{ http://wapcos.gov.in/}} for any website.   

\textit{Guideline 2.4.11 Focus Appearance (Minimum) (AA)} states that when UI components get keyboard focus then the focus indicator should be clearly visible and this indicates where a user is on the page. This goes beyond the existing 2.4.7 visible focus requirement, by defining a minimum size of the focus indicator and a minimum contrast. The level AA variant expects a contrast ratio of 3:1. 
All the websites chosen violated this guideline at least once. Majority of the websites violated this guideline at least 3 times and at most 9109\footnote{\url{https://www.mstcindia.co.in/}} times. 

\textit{Guideline 2.4.12 Focus Appearance (Enhanced) (AAA)} also focuses on appearance of elements on focus, however calling for higher level of conformance. This level AAA variant expects a contrast ratio of 4.5:1. None of the websites passed the success criterion for this guideline with cumulative violations reaching 1.6 million. 

\textit{Guideline 1.1.1 Non-text Content (A)} 
requires text alternative for information conveyed by non-text content. 
Visually impaired users can hear the alt text of the image with a screen reader. Individuals with hearing disabilities can view the text description of an audio information.  
This guideline has been violated at least once by 1996 among the 2227 websites chosen, with maximum number of violations observed for a website\footnote{\url{http://westbengalforest.gov.in/}} at 1030.

\textit{Guideline 1.3.1 Info and Relationships (A)} states that information and relationships presented through auditory or visual formatting should be preserved  when the presentation format changes. User agents can be benefited by the success criterion as user agents can adapt content according to user needs. 
1377 websites have not violated this guideline, while it has been violated for a maximum of 87 times among the websites chosen. 

\textit{Guideline 1.4.3 Contrast (Minimum) (AA)} intends to provide enough contrast between the text in its background so that it can be read by users with moderately low vision without the need for an additional contrast enhancing assistive technology. This is a level AA variant requiring a minimum contrast ratio of 1:4.5 for smaller fonts and 1:3 for larger fonts. The guideline contrast minimum (1.4.3), was violated at least once by 2168 websites, with a maximum at 5044 by any website. Violations ranging from 61 to 500 make 67.4\% of the non-zero violations. 

\textit{Guideline 1.4.6 Contrast (Enhanced) (AAA)} is a level AAA variant and requires a minimum contrast ratio of 1:7 for smaller fonts and 1:4.5 for larger fonts. Only 2.5\% of the websites do not violate this guideline. Maximum of 5305 violations were observed for this guideline. 

\textit{Guideline 1.4.4 Resize Text (AA)} states that without the help of an assistive technology, text (omitting captions and images of text) should be able to be resized up to 200 percent without loosing its functionality or content. This helps users with low vision to increase the text size and read better. As part of this guideline we tried to highlight HTML elements such as bold, italic and font and suggested not to use them as these elements will be able handled well by screen readers.   
One among the 2227 websites chosen, violated this guideline for a maximum of 1518\footnote{\url{http://www.upe.bsnl.co.in/}} times. Maximum number of websites violating this guideline fell in the violation range 1-10.  

\textit{Guideline 2.2.2 Pause, Stop, Hide (A)} requires the functionality of content being operable with a keyboard. Visually impaired people would use keyboard with a screen reader. Users with low vision can be benefited with content that is navigable through keyboard.
Nearly 37\% of the websites found to be violating this guideline. Almost all the violations referring to this guideline were ranged from 1-10, with a maximum violated by any website at 13\footnote{\url{http://techedu.rajasthan.gov.in/}}. 

\textit{Guideline 2.4.4 Link Purpose (In Contrast) (A)} states that the purpose of each link should be able to be determined from the link's text alone or from the link's text in combination with its programmatically determined accessibility name. The success criterion helps people with motion impairment in deciding whether they should follow a link and they can skip pages that they are not interested in visiting. Only 8\% of the websites found to be meeting the requirements referring to this guideline. Majority of the websites that violated this guideline fell in the range 61-500, with maximum number of violations at 1892\footnote{\url{http://sknau.ac.in/}} for any website.

\textit{Guideline 2.4.6 Headings and Labels (AA)} states that headings or labels should describe the topic or purpose. Clear headings will enable users to get information more easily. For individuals with reading disabilities or short-term memory might find descriptive headings useful. In the implementation of \textit{WAccess} we evaluated the ordering for header tags and did not verify the purpose as it requires manual investigation.     
Of the 2227 websites 2208 were found violating the guideline and number of violations are one each per website.  

\textit{Guideline 3.1.1 Language of Page (A)} requires the HTML page to have the a valid language id defined in the \textit{lang} attribute. Failing to meet this guideline criterion makes it difficult to identify the language of the page and might affect the screen reader's performance in catching the correct accent or pronunciation. Meeting the guideline requirements will help the people who use text-to-speech software. We only calculate the length of the language code in the implementation of \textit{WAccess}. Of the 2227 websites 2187 found violating the guideline and number of violations were observed to be one per website.  

\textit{Guideline 3.3.2 Labels or Instructions (A)} requires labels or instructions to be provided to the user for identifying controls in a form such that users know the expected input data. Meeting the success criterion helps people with cognitive, language and learning disabilities to enter the correct information. This guideline has been violated at least once by 1678 among the 2227 websites chosen, with maximum number of violations observed for a website at 9801.

\textit{Guideline 4.1.1 Parsing (A)} requires the HTML tags to have opening and closing tags. This guideline also does not allow duplicate id attributes. Meeting the guideline's success criterion enables assistive technologies parse the content without issues. 73\% of the websites have not violated this guideline, while maximum number of violations observed for any website is at 242\footnote{\url{https://keralataxes.gov.in/}}.

\begin{figure*}
  \centering
  \includegraphics[width = \linewidth]{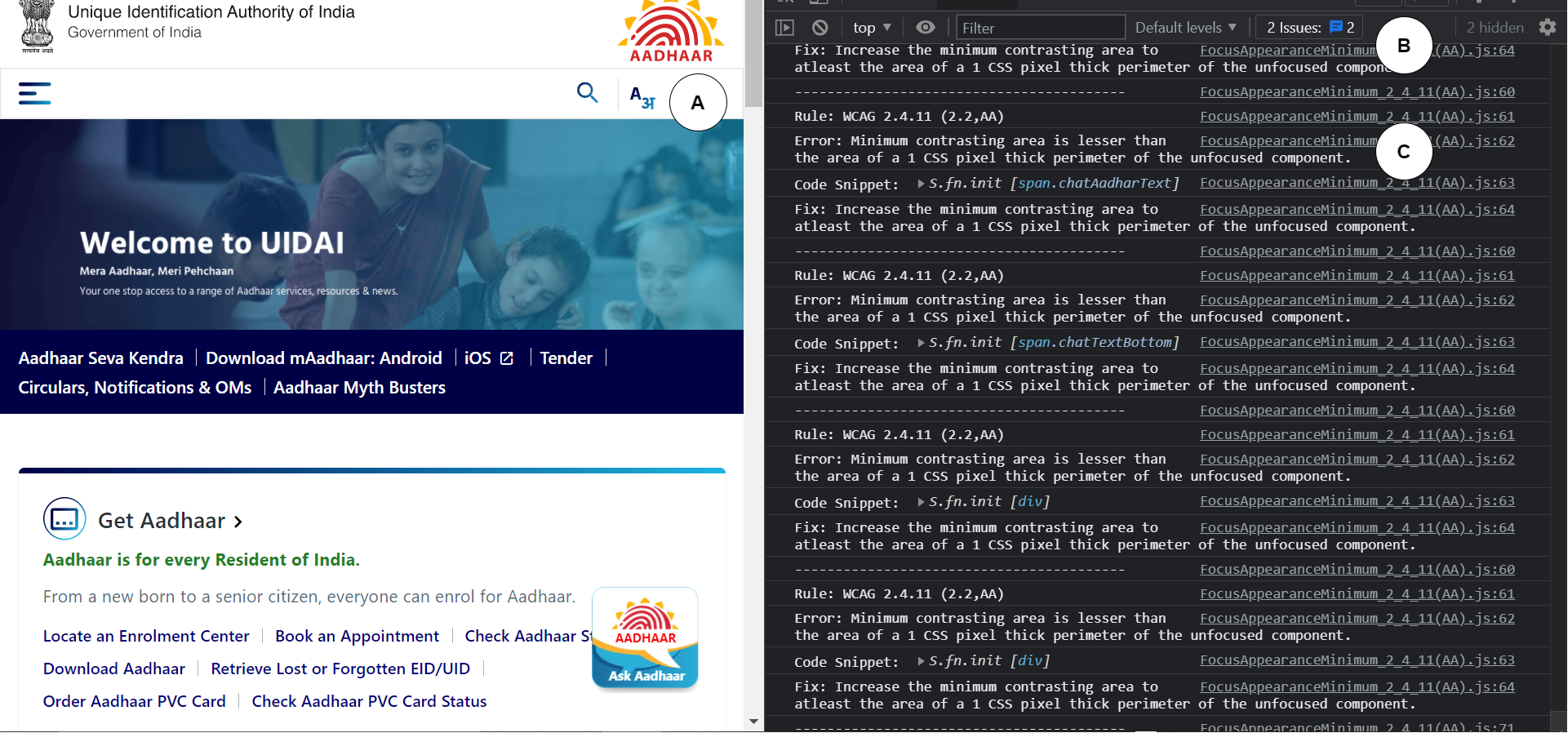}
  \caption{A snapshot depicting results of \textit{WAccess}. [A] depicts the UIDAI website which is evaluated by \textit{WAccess}. [B] is the web console highlighting the list of errors with respect to guidelines. [C] is the violation block pertaining to a guideline consisting respective WCAG guideline ID, error,  code snippet, and fix}
  \label{fig:aadhaar}
\end{figure*}

% \begin{figure*}
%   \centering
%   \includegraphics[width = \linewidth]{Waccess-cs2-cu.png}
%   \caption{Snapshots depicting results of \textit{WAc cess}. [A] depicts the Commerce website which is evaluated by \textit{WAccess}. [B] highlights the option to be selected to trigger \textit{WAccess}- Accessibility evaluation. [C] shows a list of errors identified by \textit{WAccess} based on the guidelines. [D], [E], [F], [G], [H] and [I] likewise represent violations with respect to different guidelines}
%   \label{fig:commerce}
% \end{figure*}

\subsubsection{Guidelines with zero violations}
In the evaluation of 2227 websites, the guidelines 1.4.1, 1.4.13, 2.1.1, 2.1.4, 2.3.3, 3.2.7, and 4.1.3 displayed zero violations. This does not imply that the success criterion of these guidelines has been completely achieved as the implementation of \textit{WAccess} might not be complete. For example, in guideline 2.1.1 that checks the operability through keyboard, checking the effect of an image on receiving focus from keyboard on the rest of layout would be possible to be identified only by manually assessing the visual outlook of the web page and cannot be programmatically verified. Hence, zero violations for 2.1.1 guideline for a specific website does not claim its conformance.   

% To illustrate this, for Guideline 2.1.1, \textit{WAccess} checks if a HTML element is keyboard accessible by finding whether element's defined \textit{onblur} function is missing the defined \textit{onmouseout} function. In this case, no website among 2227 websites might have defined \textit{onmouseout} and \textit{onblur} functions, which do not trigger the violation as per the rules framed for the guideline inside \textit{WAccess}.    

\subsection{Illustration of \textit{WAccess}}
We demonstrate the usage of \textit{WAccess} by navigating across two Indian government websites, UIDAI (Fig. \ref{fig:aadhaar}[A]), and Commerce website (Fig. \ref{fig:commerce}[A]). The UIDAI website contains unique identification details of all citizens in the country and is used by billions of Indian population, and the Commerce website contains services and merchandise with respect to foreign trade and public sector.
% To check the accessibility of the plugin, we selected the "Accessibility Test" option from the drop-down menu, as shown in Fig. \ref{fig:aadhaar}[B] and Fig. \ref{fig:commerce} [B] , obtained by right-clicking the mouse. 
Fig. \ref{fig:aadhaar}[B] and Fig. \ref{fig:commerce} [B] displays a list of deviations from the accessibility guidelines as errors identified by \textit{WAccess} with respect to the defined guidelines, the code snippet that caused the violation, and a suggested fix. 
These errors are presented with respect to each guideline as represented in Fig. \ref{fig:aadhaar}[C], and Fig. \ref{fig:commerce}[C].
We observed that \textit{WAccess} could list out guidelines that are not being followed by a website from the 13 WCAG 2.0, 9 WCAG 2.1 and 7 WCAG 2.2 guidelines considered in its design. Number of violations observed for both of these websites are presented in Table \ref{tab:case_study}

\subsubsection{UIDAI website:}
Through \textit{WAccess,} we found 10,339 guideline violations on this website. Nearly 89\% of the violations attribute to the guidelines 1.4.3, 1.4.6, 2.5.8, 2.4.11 and 2.4.12. Less than 10 violations were observed for the guidelines 1.3.1, 1.3.5, 1.4.11, 2.4.6, 2.4.13, 3.1.1 and 3.3.7.   
% With respect to accessibility standards guidelines 2.5.3 and 3.3.7 conform to level A, guidelines 1.3.5, 1.4.11, 2.4.11 and 2.5.7 to level AA, and guidelines 1.3.6, 2.5.5, and 2.4.12 conform to level AAA.
Guidelines conforming to conformance level AA, took a significant share in the number of violations (about 59\%), while for A, the number of violations were observed to be lesser (only around 5.2\%). Guidelines referring to minimum conformance level AAA, formed 36.4\% of the total violations. 
\subsubsection{Commerce website:}
\begin{figure*}
  \centering
  \includegraphics[width = \linewidth]{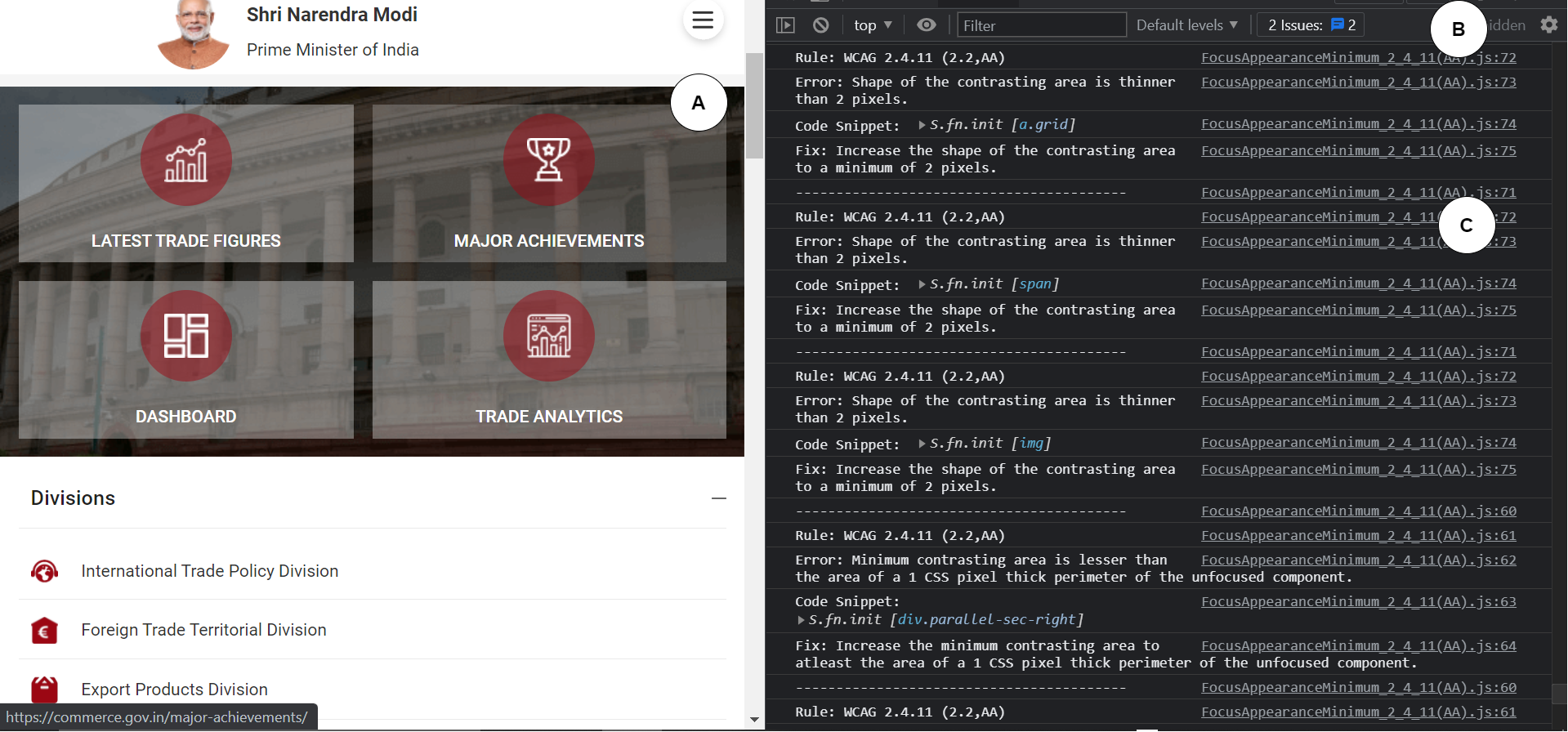}
  \caption{A snapshot depicting results of \textit{WAccess}. [A] depicts the Commerce website which is evaluated by \textit{WAccess}. [B] is the web console highlighting the list of errors with respect to guidelines. [C] is the violation block pertaining to a guideline consisting respective WCAG guideline ID, error,  code snippet, and fix}
  \label{fig:commerce}
\end{figure*}

Demonstration of using \textit{WAccess} for the commerce website is depicted in Fig. \ref{fig:commerce}. Through \textit{WAccess,} we found 10,059 guideline violations on this website. Violations referring to guidelines 1.4.6, 2.4.11, 2.4.12 and 2.5.8 constitute a share of 74\% of the total violations. 
% About 14.5\% of the violations failed to meet the 2.5.7 guideline requirements.
Less than 10 violations were observed for the guidelines 1.3.1, 1.3.5, 1.4.4, 2.4.6, 3.1.1 and 3.3.7. Guidelines conforming to conformance level AA, took a significant share in the number of violations (about 62\%), while for the conformance level A, the number of violations were observed to be the least (only around 8.46\%). Guidelines referring to minimum conformance level AAA, formed 30\% of the total violations. 

\section{Discussion and Limitations}
\label{sec:discussion}
In this paper, we presented \textit{WAccess}, a tool for checking web accessibility, based on WCAG 2.0, 2.1 and 2.2 guidelines. 
\textit{WAccess} evaluates accessibility with respect to 13 WCAG 2.0, 9 WCAG 2.1 and 7 WCAG 2.2 guidelines. 

Though WCAG 2.0 and WCAG 2.1 comprise more number of guidelines, some of them require human intervention, restricting the scope for automated evaluation of the websites. Also, some of the selected guidelines contain rules (success criteria) of which few might require manual intervention. In order to improve the coverage of \textit{WAccess}, we selected only those human-independent parts of success criteria of the guidelines, that can be realised by automated analysis. As a result, if a specific guideline is satisfied, only a part of the guideline could be satisfied (partially satisfied) and it can not be assured that the whole guideline is satisfied, thus indicating the scope of false positives. An approach that includes artificial intelligence aspects to simulate manual analysis could be explored and integrated with \textit{WAccess} to address this concern.
However, there exists no false negatives, i.e., if a website is marked to be violating a guideline, it is assured not to be satisfying the specific guideline.

The manual analysis to assess the correctness of \textit{WAccess} was conducted only on two randomly selected websites of the 2227 websites considered for automated analysis. The correctness of the violations displayed in the console of these websites based on the rule-based approach followed by \textit{WAccess} was manually validated. While care has been taken to incorporate the success criteria of the guidelines in the rule-based approach, thus validating the correctness of \textit{WAccess}, a large-scale manual analysis could be performed to provide stronger insights on its correctness.

The current version of \textit{WAccess} is an extension to the browser and can be incorporated into automated scripts to analyse large number of websites. However, a command-line interface version of the tool could ease the automation task further.
% Hence, in the implementation of a guideline, if an extent of the success criterion needs human effort, in that case we resorted to implementing the human-independent part of the success criterion. 
% \color{red}
\\
\\

\fbox{
\begin{minipage}{45em}
\textit{\textbf{Key findings:}}
\begin{itemize}
    \item Selected guidelines implemented in \textit{WAccess} comprise 14 guidelines of level A conformance, 10 guidelines of level AA conformance, and 5 guidelines of level AAA conformance. 
    \item No website among the 2227 chosen, complies with all of the three conformance levels.
    \item Cumulatively, guidelines conforming to minimum level (A) make nearly 54.5\% of total 6.1 million violations. 
    \item Level AAA violations take the next huge share among the observed violations at 2.2 million and Level AA constitute about 10\% of total violations.  
    \item We have highlighted the accessibility violations for 2227 Indian government websites with \textit{WAccess}, and observed that violations pertaining to considered WCAG 2.2 guidelines make around 4.57 million which is a significant share of the total 6.1 million violations. Among the considered guidelines, WCAG 2.0 constitutes 25\% and WCAG 2.1 only 2.8\% of 6.1 million violations.  
    \item However, this result is based only on the 29 considered guidelines from WCAG 2.0, 2.1 and 2.2 series and  might differ if all the WCAG 2.0, 2.1 and 2.2 guidelines are considered for evaluating the websites.
\end{itemize}
\end{minipage}
}
\newline

% \newline We have randomly chosen two websites and randomly evaluated each guideline's authenticity and manually verified the rule based approach implemented for each guideline. We found no false positives or false negatives for the random investigation and found the accuracy to be 100 percent.    

\section{Conclusion and Future Work}
\label{sec:conclusion}
In this paper, we presented \textit{WAccess}, as an open source tool to assess the web accessibility of websites based on WCAG 2.0, 2.1 and 2.2 guidelines. Though there are multiple tools available to evaluate websites against WCAG 2.1 guidelines, these tools do not support automated evaluation of large number of websites, and are not open source. Further, there do not exist any tools to check the conformance of a website against WCAG 2.2 guidelines.

\textit{WAccess} is a browser extension based on a total of 29 WCAG guidelines, 13 from WCAG 2.0, 9 from WCAG 2.1 and 7 from WCAG 2.2, and supports large scale accessibility evaluation. We used \textit{WAccess} to automatically detect accessibility violations in 2227 Government of India websites. The results of the evaluation showed the deviations of each website with respect to the 29 guidelines being considered. These deviations are displayed as errors in each web page's browser-console, along with the code snippet that caused the deviation and a possible fix to rectify the deviation.    

\textit{WAccess} can further be explored to include a broader scope of guidelines by avoiding human intervention through the implementation of advanced techniques in Artificial Intelligence and Machine Learning domains. We also plan to enhance the existing version of \textit{WAccess} by improving the user interface of the tool and by employing better technologies for the development of the tool.
The current browser extension based version of \textit{WAccess} could be extended as a command-line interface version to further ease the task of automated analysis.

\textit{WAccess} currently suggests fixes to the webpage based on the violations against WCAG 2.0, 2.1 and 2.2 guidelines. It can be further improved to support automated or semi-automated refactoring of the websites during website development, thus, consequently helping web developers in abiding by the accessibility guidelines, towards making the websites accessible to everyone. Other forms of \textit{WAccess} tool, such as open API, webpage, and so on, can also be developed to support a broader range of audience, and a wider range of studies, aimed to analyze the accessibility of websites.

\textit{WAccess} could be extended to the newly drafted WCAG 3.0 guidelines in the future, and can also be extended to support other domain-specific accessibility guidelines, such as GIGW, a set of guidelines for government websites in the Indian context.

\begin{acks}
We thank our undergraduate students, Krishnendu Sudheesh, Nikhil, Soham and Ishaan for helping us with a part of the development of \textit{WAccess}.
\end{acks}

\bibliographystyle{Reference-Format/ACM-Reference-Format}
\bibliography{waccess-references}

\end{document}